\documentclass[10pt]{article}
\usepackage{graphicx}
\usepackage{amsmath}
\usepackage{amssymb}
\usepackage{caption2}
\usepackage[figuresright]{rotating}
\begin{document}
\begin{center}
{\large {\bf \sc{  Analysis of the pseudoscalar hidden-charm tetraquark  states with  the QCD sum rules }}} \\[2mm]
Zhi-Gang  Wang \footnote{E-mail: zgwang@aliyun.com.  }, Qi Xin     \\
 Department of Physics, North China Electric Power University, Baoding 071003, P. R. China
\end{center}

\begin{abstract}
In this work,  we take all the color-antitriplet diquarks, such as the scalar, pseudoscalar,  vector,  axialvector and tensor diquarks,  as the basic constituents to construct the local pseudoscalar  four-quark currents  without importing the explicit P-waves to implement the negative-parity,  and investigate the mass spectroscopy of the pseudoscalar  hidden-charm tetraquark states without strange, with strange and with hidden-strange in the framework of  the QCD sum rules in  a consistent and comprehensive   way. We obtain the lowest mass $4.56\pm0.08\,\rm{GeV}$ for the pseudoscalar tetraquark state with the symbolic quark constituents  $c\bar{c}u\bar{d}$, which is much larger than the experimental value   $4239\pm18{}^{+45}_{-10}\,\rm{MeV}$ extracted by the LHCb collaboration.
 \end{abstract}

 PACS number: 12.39.Mk, 12.38.Lg

Key words: Tetraquark  state, QCD sum rules

\section{Introduction}
In recent years, a number of  exotic charmonium-like states have been reported  by the  BaBar, Belle, BESIII, CDF, CMS, D0, LHCb collaborations \cite{PDG}, such as the $X(3860)$, $X(3872)$, $Z_c(3900)$, $X(3915)$, $X(3940)$, $Z_{cs}(3985)$, etc, which have hidden-charm and therefore are charmonium-like, the traditional  quark models are facing great challenges,  we have to resort to additional quark and gluon constituents beyond the naive quark-antiquark pairs to probe  their properties, especially for those  states with non-zero electric-charge, such as the $Z^\pm_c(3900)$, $Z^\pm_c(4020)$, $Z^\pm_c(4430)$, etc. Several  states  lie slightly above or below the meson-antimeson or meson-meson  thresholds provocatively, which maybe indicate that they are very good candidates for the
hadronic molecules or threshold effects, but the others do
not lie near the two-meson thresholds, why? Do they have absolutely different inner structures in nature?   The theoretical physicists  have tentatively suggested several possible assignments of those $X$, $Y$ and $Z$ states, such as the diquark-antidiquark type tetraquark states, hadronic molecular states (or color-singlet-color-singlet type tetraquark states), hadroquarkonium, dynamically generated resonances (another type molecular states),
kinematical effects, cusp effects,  virtual states, etc.

In the charm sector, if we focus on (or employ) the scenario of diquark-antidiquark type tetraquark states, up to now, only experimental candidates for the hidden-charm tetraquark states with the spin-parity $J^P=0^+$, $1^+$, $1^-$ have been reported, no experimental candidates for the spin-parity $J^P=0^-$ and $2^+$ hidden-charm tetraquark states have ever been reported  yet \cite{PDG}.
In 2014, the LHCb collaboration performed a four-dimensional fit of the $B^0\to\psi'\pi^-K^+$ decay amplitude, and provided the first independent confirmation of the
existence of the $Z_c(4430)$  in the $\psi^\prime \pi^-$ mass spectrum and established its spin-parity to be $J^P=1^+$ \cite{LHCb-Zc4430-Rc4240}, which excludes the possibility of  assigning the $Z_c(4430)$ as the $D^*D_1$ molecular state with the spin-parity $J^P=0^-$ \cite{Nielsen-Zc4430}, although it lies near the $D^*D_1$ threshold. Furthermore, the LHCb collaboration observed a weak evidence  for an additional resonance, the $Z_{c}(4240)$,  in the $\psi^\prime \pi^-$ mass spectrum with the preferred   spin-parity $J^P=0^-$, and the Breit-Wigner mass $M_{Z}=4239\pm18{}^{+45}_{-10}\,\rm{MeV}$ and width $\Gamma_Z=220\pm47\,{}^{+108}_{-\phantom{0}74}\,\rm{MeV}$, respectively with large uncertainties \cite{LHCb-Zc4430-Rc4240}.  If the $Z_{c}(4240)$ is confirmed by further experimental research/detection   in the future, it is an excellent candidate for the pseudoscalar hidden-charm tetraquark state with the spin-parity-charge-conjugation $J^{PC}=0^{--}$.

The attractive interactions inferred from one-gluon exchange support making diquark correlations in the color-antitriplet $\bar{\bf 3}_c$ come into being \cite{One-gluon-1,One-gluon-2}. The diquark operators $\varepsilon^{ijk}q^{T}_j C\Gamma q^{\prime}_k$ in the $\bar{\bf 3}_c$ have  five spinor  structures, where the color indexes $i$, $j$, $k=1$, $2$, $3$,  the Dirac matrixes $C\Gamma=C\gamma_5$, $C$, $C\gamma_\mu \gamma_5$,  $C\gamma_\mu $ and $C\sigma_{\mu\nu}$ correspond to the scalar, pseudoscalar, vector, axialvector  and  tensor diquark states, respectively.
The $C\gamma_5$-type and $C\gamma_\mu$-type diquark states have the spin-parity $J^P=0^+$ and $1^+$, respectively, the corresponding  $C$-type and $C\gamma_\mu\gamma_5$-type diquark states have the spin-parity $J^P=0^-$ and $1^-$, respectively, the  P-wave effects   are implicitly embodied in the underlined  Dirac gamma matrix, $\gamma_5$, in the  $C\gamma_5 \underline{\gamma_5} $ and $C\gamma_\mu \underline{\gamma_5} $, which changes their  parity.  The $C\sigma_{\mu\nu}$-type and $C\sigma_{\mu\nu}\gamma_5$-type  diquark states have both the spin-parity $J^P=1^+$ and $1^-$ components,  the P-waves  effects   are implicitly embodied in the negative parity. In this work, we take the elementary building blocks  $C\gamma_5$, $C$, $C\gamma_\mu \gamma_5$,  $C\gamma_\mu $, $C\sigma_{\mu\nu}$ and $C\sigma_{\mu\nu}\gamma_5$ diquark operators  to construct the  diquark-antidiquark type local four-quark currents  without importing the explicit P-waves, which are usually presented by the derivative  $\stackrel{\leftrightarrow}{\partial}_\mu=\stackrel{\rightarrow}{\partial}_\mu-\stackrel{\leftarrow}{\partial}_\mu$, and investigate the mass spectroscopy of the pseudoscalar hidden-charm  tetraquark states in the framework of  the QCD sum rules, and explore the possible assignment of the $Z_c(4240)$.

The QCD sum rules method  is  a vigorous and  powerful theoretical tool in probing  the exotic $X$, $Y$ and $Z$ states, there have been several  possible assignments of the  $X$, $Y$ and $Z$ states, such as the diquark-antidiquark type tetraquark states, the tetraquark molecular states, the $c\bar{c}$-tetraquark mixing states, according to the analysis via the QCD sum rules  \cite{MNielsen-review-1812}.
The predictions depend to a great extent   on the particular schemes in which the input parameters are accepted/adopted  at the QCD side, for example, even in the same diquark-antidiquark type tetraquark scenario, the same  quark currents can  result in quite different predictions therefore quite different  assignments of the $Y$ states  \cite{Nielsen-4260-4460,ChenZhu-Vector-Axial,WangY4360Y4660-1803}. A comprehensive and consistent investigation of all the scalar, pseudoscalar, vector, axialvector and tensor hidden-charm tetraquark (molecular) states  with the same input parameters and same treatments of the operator product expansion is necessary to avoid possible  biased predictions.

In Refs.\cite{WZG-HC-spectrum-PRD,WZG-Zcs-spectrum-CPC,WZG-Vector-spectrum-NPB} (\cite{WZG-Axial-cc-APPB,WZG-cc-spectrum-EPJC}), we construct the diquark-antidiquark type four-quark currents without importing the explicit P-waves to investigate   the  mass spectroscopy of the ground state hidden-charm (doubly-charmed) tetraquark states with the $J^{PC}=0^{++}$, $1^{++}$, $1^{+-}$, $1^{--}$, $1^{-+}$ and $2^{++}$ ($0^{++}$, $1^{++}$, $1^{+-}$ and $2^{++}$) in the framework of
the QCD sum rules in a comprehensive and consistent way, and revisit the assignments of all the observed  $X$, $Y$, $Z$ states in the  scenario of tetraquark  states, and make a series of predictions which can  be compared with  the experimental data in the future to illustrate (demonstrate) the nature of the tetraquark states and their inner quark-gluon structures.  The predicted masses of  the diquark-antidiquark  type  axialvector $cc\bar{u}\bar{d}$ tetraquark states  are  $3.90\pm0.09\,\rm{GeV}$ \cite{WZG-Axial-cc-APPB,WZG-cc-spectrum-EPJC}, which are in very good  agreement with the experimental value from the LHCb collaboration later \cite{LHCb-Tcc,LHCb-Tcc-detail}.
In Refs.\cite{Vector-Tetra-WZG-P-wave-1,Vector-Tetra-WZG-P-wave}, we import an explicit P-wave between the diquark and antidiquark building blocks to implement the negative-parity to construct the local four-quark currents to explore   mass spectroscopy of the ground state hidden-charm tetraquark states with the $J^{PC}=1^{--}$ in an  systematic way, and obtain the lowest vector tetraquark states up to today.

In Ref.\cite{WZG-mole-spectrum-IJMPA}, we construct  the color-singlet-color-singlet type (or meson-meson type)  local four-quark  currents to  investigate   the mass spectroscopy of the  hidden-charm  tetraquark molecular states with the $J^{PC}=0^{++}$, $1^{++}$, $1^{+-}$ and $2^{++}$ in a comprehensive and consistent way and make possible assignments of all the $X$, $Y$ and $Z$ states in a different scheme, and observe that the scenario of tetraquark molecule states can accommodate much less exotic states than that of the diquark-antidiquark type tetraquark  states.

In all the works \cite{WZG-HC-spectrum-PRD,WZG-Zcs-spectrum-CPC,WZG-Vector-spectrum-NPB,WZG-Axial-cc-APPB,WZG-cc-spectrum-EPJC,Vector-Tetra-WZG-P-wave-1,Vector-Tetra-WZG-P-wave,WZG-mole-spectrum-IJMPA}, we resort to our unique benchmark, the energy scale formula  $\mu=\sqrt{M^2_{X/Y/Z}-(2{\mathbb{M}}_c)^2}$ or its modifications with the effective $c$-quark mass ${\mathbb{M}}_c$, which has the universal value,  to acquire the best  energy scales characterizing the QCD spectral densities via trial and error.  The energy scale formula plays an essential role in increasing  the pole contributions and in making  the convergent behaviors of the operator product expansion much better \cite{Wang-tetra-formula}, and it is a unique feature of our works.

In  this work, we extend our previous works \cite{WZG-HC-spectrum-PRD,WZG-Zcs-spectrum-CPC,WZG-Vector-spectrum-NPB,WZG-Axial-cc-APPB,WZG-cc-spectrum-EPJC,Vector-Tetra-WZG-P-wave-1,Vector-Tetra-WZG-P-wave,WZG-mole-spectrum-IJMPA} to explore   the mass spectroscopy of the pseudoscalar hidden-charm tetraquark states and make great effects to obtain comprehensive investigations on the hidden-charm tetraquark states in a consistent way. We take the elementary constituents $C\gamma_5$, $C$, $C\gamma_\mu \gamma_5$,  $C\gamma_\mu $, $C\sigma_{\mu\nu}$ and $C\sigma_{\mu\nu}\gamma_5$ (anti)diquark operators    to construct the local  four-quark pseudoscalar   currents without resorting to the explicit P-wave. While in the dynamical (di)quark models and constituent (di)quark models, we always  take only the scalar and axialvector diquarks, the most stable diquarks, as the basic building blocks to investigate all the scalar, pseudoscalar, vector, axialvector and tensor hidden-charm tetraquark states, and explicit P-waves between the diquark and antidiquark constituents are needed to acquire  the pseudoscalar and vector tetraquark states \cite{Lebed-PRD-2020,Ferretti-PRD-2018,QFLu-PRD-2016}. The tetraquark spectroscopy obtained  in Refs.\cite{Lebed-PRD-2020,Ferretti-PRD-2018,QFLu-PRD-2016} differ from the present work significantly. We investigate  the mass spectroscopy of the pseudoscalar hidden-charm tetraquark states in the framework of  the QCD sum rules comprehensively, and take account of the light flavor $SU(3)$ breaking effects, such as the quark masses and vacuum condensates,    and resort to the modified energy scale formula  $\mu=\sqrt{M^2_{X/Y/Z}-(2{\mathbb{M}}_c)^2}-k\, m_s(\mu)$ with $k=0$, $1$, $2$ to get  the optimal  energy scales characterizing  the QCD spectral densities. In Ref.\cite{Chen-Zhu-PRD-2010}, Chen and Zhu  study the mass spectroscopy of the pseudoscalar hidden-charm tetraquark states in the framework of  the QCD sum rules, their interpolating currents and particular scheme in treating the operator product expansion and input parameters at the QCD side   differ from this work remarkably.

The article is arranged as follows:  we obtain   the QCD sum rules for the masses and pole residues  of  the pseudoscalar  hidden-charm tetraquark states in section 2; in section 3, we   exhibit  the numerical results and discussions; section 4 is reserved for our conclusion.

\section{QCD sum rules for  the  pseudoscalar  hidden-charm  tetraquark states}
According to the routine of the QCD sum rules, we  write down  the two-point correlation functions  $\Pi(p)$ firstly,
\begin{eqnarray}\label{CF-Pi}
\Pi(p)&=&i\int d^4x e^{ip \cdot x} \langle0|T\Big\{J(x)J^{\dagger}(0)\Big\}|0\rangle \, ,
\end{eqnarray}
where the local four-quark currents,
\begin{eqnarray}
J(x)&=&J^{+}_{AV}(x)\, ,\,\, J^{-}_{AV}(x)\, , \,\,J_{PS}^{+}(x)\, , \,\, J_{PS}^{-}(x)\, ,\,\,
J_{TT}^{+}(x)\, ,\,\,J_{TT}^{-}(x)\, ,
\end{eqnarray}
\begin{eqnarray}
J_{AV}^{+}(x)&=&\frac{\varepsilon^{ijk}\varepsilon^{imn}}{\sqrt{2}}\Big[q^{Tj}(x)C\gamma_{\mu}c^k(x) \bar{q}^{\prime m}(x)\gamma_5\gamma^\mu C \bar{c}^{Tn}(x)-q^{Tj}(x)C\gamma_\mu\gamma_5 c^k(x)\bar{q}^{\prime m}(x)\gamma^{\mu}C \bar{c}^{Tn}(x) \Big] \, ,\nonumber\\
J_{AV}^{-}(x)&=&\frac{\varepsilon^{ijk}\varepsilon^{imn}}{\sqrt{2}}\Big[q^{Tj}(x)C\gamma_{\mu}c^k(x) \bar{q}^{\prime m}(x)\gamma_5\gamma^\mu C \bar{c}^{Tn}(x)+q^{Tj}(x)C\gamma_\mu\gamma_5 c^k(x)\bar{q}^{\prime m}(x)\gamma^{\mu}C \bar{c}^{Tn}(x) \Big] \, ,\nonumber\\
J_{PS}^{+}(x)&=&\frac{\varepsilon^{ijk}\varepsilon^{imn}}{\sqrt{2}}\Big[q^{Tj}(x)C c^k(x) \bar{q}^{\prime m}(x)\gamma_5 C \bar{c}^{Tn}(x)+q^{Tj}(x)C\gamma_5 c^k(x)\bar{q}^{\prime m}(x)C \bar{c}^{Tn}(x) \Big] \, ,\nonumber\\
J_{PS}^{-}(x)&=&\frac{\varepsilon^{ijk}\varepsilon^{imn}}{\sqrt{2}}\Big[q^{Tj}(x)C c^k(x) \bar{q}^{\prime m}(x)\gamma_5 C \bar{c}^{Tn}(x)-q^{Tj}(x)C\gamma_5 c^k(x)\bar{q}^{\prime m}(x)C \bar{c}^{Tn}(x) \Big] \, ,\nonumber\\
J_{TT}^{+}(x)&=&\frac{\varepsilon^{ijk}\varepsilon^{imn}}{\sqrt{2}}\Big[q^{Tj}(x)C\sigma_{\mu\nu}c^k(x) \bar{q}^{\prime m}(x)\gamma_5\sigma^{\mu\nu} C \bar{c}^{Tn}(x)+q^{Tj}(x)C\sigma_{\mu\nu}\gamma_5 c^k(x)\bar{q}^{\prime m}(x)\sigma^{\mu\nu}C \bar{c}^{Tn}(x) \Big] \, ,\nonumber\\
J_{TT}^{-}(x)&=&\frac{\varepsilon^{ijk}\varepsilon^{imn}}{\sqrt{2}}\Big[q^{Tj}(x)C\sigma_{\mu\nu}c^k(x) \bar{q}^{\prime m}(x)\gamma_5\sigma^{\mu\nu} C \bar{c}^{Tn}(x)-q^{Tj}(x)C\sigma_{\mu\nu}\gamma_5 c^k(x)\bar{q}^{\prime m}(x)\sigma^{\mu\nu}C \bar{c}^{Tn}(x) \Big] \, ,\nonumber\\
\end{eqnarray}
 with $q$, $q^\prime=u$, $d$, $s$, the $i$, $j$, $k$, $m$, $n$ are  color indexes,  the $C$ is the charge-conjugation matrix, the superscripts $\pm$ symbolize the positive  and negative charge-conjugation, respectively, the subscripts $P$, $S$, $V$,  $A$ and $T$ stand for  the pseudoscalar, scalar, vector, axialvector and tensor diquark (and antidiquark) operators, respectively.
  Under  parity transform $\widehat{P}$, the four-quark currents $J(x)$ have the  property,
\begin{eqnarray}\label{J-parity}
\widehat{P} J(x)\widehat{P}^{-1}&=&-J(\tilde{x}) \, ,
\end{eqnarray}
which warrants that the Lorentz scalar currents $J(x)$ have the negative parity, therefore they are pseudoscalar currents.
Under  charge-conjugation transform $\widehat{C}$, the four-quark currents  $J(x)$  have the properties,
\begin{eqnarray}
\widehat{C}J^{\pm}(x)\widehat{C}^{-1}&=&\pm J^{\pm}(x)\mid_{q\leftrightarrow q^\prime}  \,   ,
\end{eqnarray}
which warrants that  we can distinguish the positive and negative charge-conjugations unambiguously. By the way, we can prove that the current $J_{TT}^{-}(x)=0$ through performing the Fierz-transformation.

The currents $J^{+}_{AV}(x)$, $\,J_{PS}^{+}(x)$ and $J_{TT}^{+}(x)$ have the same quantum numbers $J^{PC}=0^{-+}$, while the currents $J^{-}_{AV}(x)$ and $J_{PS}^{-}(x)$ have the same quantum numbers $J^{PC}=0^{--}$. The currents having the same quantum numbers could mix with each other under re-normalization, we have to import mixing matrixes (or transformation matrixes) $\mathcal{M}$ to obtain diagonal currents $\tilde{J}(x)$ under re-normalization,
\begin{eqnarray}
\tilde{J}&=&\mathcal{M}\, J \, , \nonumber\\
\gamma_{\tilde{J}}&=&\mathcal{M}\, \gamma_J \, \mathcal{M}^{-1}\, ,
\end{eqnarray}
where the $\gamma_J$ are the  anomalous dimension matrixes of the current operators $J(x)$, and the $\gamma_{\tilde{J}}$ are the  diagonal anomalous dimension matrixes. In the present case, the matrixes $\gamma_J$ are $3\times3$ or $2\times2$ matrixes  in the case of $J(x)= J^{+}_{AV}(x)$, $\,J_{PS}^{+}(x)$, $J_{TT}^{+}(x)$ or $J(x)=J^{-}_{AV}(x)$, $J_{PS}^{-}(x)$, respectively.  In general, the matrixes $\gamma_J$ can be expanded in terms of the strong fine structure constant $\alpha_s=\frac{g^2_s}{4\pi}$,
\begin{eqnarray}
\gamma_J&=&C_{\gamma_J,1}\frac{\alpha_s}{4\pi}+C_{\gamma_J,2}\left(\frac{\alpha_s}{4\pi}\right)^2+\cdots\, ,
\end{eqnarray}
where the $C_{\gamma_J,1}$ and $C_{\gamma_J,2}$ are the coefficients corresponding to the next-to-leading-order and next-to-next-to-leading-order radiative corrections, respectively.
If we  choose the diagonal currents $\tilde{J}(x)$,  then the current operators $\tilde{J}(x)$ do not mix under re-normalization, and are expected to couple potentially to the physical pseudoscalar tetraquark states, as the physical masses are invariant under re-normalization, they are determined by experimental detections.

Generally speaking,  a physical pseudoscalar tetraquark state, just like other hadron states, maybe have several Fock components, we can choose any current with the same quark structure as one of the Fock components to interpolate this tetraquark state due to the non-vanishing current-hadron coupling constant. Under re-normalization,
there are new components induced in this  special current operator, accordingly, we have to import new Fock components of the orders $\frac{\alpha_s}{4\pi}$, $\left(\frac{\alpha_s}{4\pi}\right)^2$, etc to match with the updated  current operator. In this aspect, the diagonalized current operators $\tilde{J}(x)$ are preferred, however, at the present time, we cannot acquire the  mixing matrixes $\mathcal{M}$ without calculating the  anomalous dimension matrixes $\gamma_J$, this maybe our next work.

Now go back to Eq.\eqref{CF-Pi}, at the hadron  side, we  insert  a complete   set of  intermediate  hadronic states,
such as the tetraquark states, two-meson scattering states, continuum states, etc,  having
the same quantum numbers  as the  current operators  $J(x)$    into the
correlation functions  $\Pi(p)$   to obtain the hadronic spectral representation
\cite{SVZ79-1,SVZ79-2,Reinders85}, and distinguish  the contributions of the lowest pseudoscalar  hidden-charm tetraquark states without strange, with strange and with hidden-strange, respectively,
\begin{eqnarray}
\Pi(p)&=&\frac{\lambda_{Z}^2}{M_{Z}^2-p^2}+\cdots \, ,
\end{eqnarray}
where the pole residues $\lambda_Z$ are defined by  $ \langle 0|J(0)|Z_c(p)\rangle =\lambda_{Z}$.
In the isospin limit $m_u=m_d$, the four-quark currents with the  symbolic quark  structures,
 \begin{eqnarray}
 \bar{c}c\bar{d}u, \, \, \bar{c}c\bar{u}d, \, \, \bar{c}c\frac{\bar{u}u-\bar{d}d}{\sqrt{2}}, \, \, \bar{c}c\frac{\bar{u}u+\bar{d}d}{\sqrt{2}}\, ,
 \end{eqnarray}
 couple potentially  to the pseudoscalar  tetraquark states with  degenerated  masses, and they result in the same  QCD sum rules as a matter of fact. On the other hand, the four-quark currents with the  symbolic quark structures,
 \begin{eqnarray}
 \bar{c}c\bar{u}s, \, \, \bar{c}c\bar{d}s, \, \, \bar{c}c\bar{s}u, \, \, \bar{c}c\bar{s}d\, ,
 \end{eqnarray}
couple also potentially  to the pseudoscalar   tetraquark states with degenerated  masses according to the isospin symmetry. Therefore, we will not distinguish the $u$ and $d$ quarks.

At the QCD side, we compute the vacuum condensates $\langle\bar{q}q\rangle$, $\langle\frac{\alpha_{s}GG}{\pi}\rangle$, $\langle\bar{q}g_{s}\sigma Gq\rangle$, $\langle\bar{q}q\rangle^2$, $\langle\bar{q}q\rangle \langle\frac{\alpha_{s}GG}{\pi}\rangle$,  $\langle\bar{q}q\rangle  \langle\bar{q}g_{s}\sigma Gq\rangle$,
$\langle\bar{q}g_{s}\sigma Gq\rangle^2$ and $\langle\bar{q}q\rangle^2 \langle\frac{\alpha_{s}GG}{\pi}\rangle$ with $q=u$, $d$ or $s$, which are vacuum expectations of the quark-gluon operators of the order $\mathcal{O}(\alpha_s^k)$ with $k\leq 1$ \cite{Wang-tetra-formula,WangHuangtao-2014-PRD,WZG-HT-EPJC-2891}, furthermore, we take the light flavor $SU(3)$ mass-breaking effects into consideration  by computing  the terms of the order $\mathcal{O}(m_s)$. In calculations, we adopt vacuum saturation for the higher dimensional vacuum condensates, for detailed discussions of this subject, one can consult Ref.\cite{Vacuum-WZG}.
Then we acquire  the spectral representation of the correlation functions $\Pi(p)$ through dispersion relation. At the end, we match  the hadron side with the QCD  side of the  $\Pi(p)$ below the continuum thresholds   $s_0$, and accomplish the Borel transform by taking the large  squared  Euclidean momentum $P^2=-p^2$ as the variable
  to get the  QCD sum rules:
\begin{eqnarray}\label{QCDSR}
\lambda^2_{Z}\, \exp\left(-\frac{M^2_{Z}}{T^2}\right)= \int_{4m_c^2}^{s_0} ds\, \rho_{QCD}(s) \, \exp\left(-\frac{s}{T^2}\right) \, ,
\end{eqnarray}
where we ignore the cumbersome analytical expressions of the QCD spectral densities $\rho_{QCD}(s)$ to save the layout of printed sheets.

We differentiate  Eq.\eqref{QCDSR} with respect to the inversed  Borel parameter   $\tau=\frac{1}{T^2}$,  and get the QCD sum rules for
 the masses of the  pseudoscalar  hidden-charm tetraquark states without strange, with strange and with hidden-strange through a fraction,
 \begin{eqnarray}\label{mass-QCDSR}
 M^2_{Z}&=& -\frac{\int_{4m_c^2}^{s_0} ds\frac{d}{d \tau}\rho_{QCD}(s)\exp\left(-\tau s \right)}{\int_{4m_c^2}^{s_0} ds \rho_{QCD}(s)\exp\left(-\tau s\right)}\, .
\end{eqnarray}

\section{Numerical results and discussions}
At first, we write down the energy-scale dependence of  the quark masses and vacuum condensates,
\begin{eqnarray}
\langle\bar{q}q \rangle(\mu)&=&\langle\bar{q}q \rangle({\rm 1GeV})\left[\frac{\alpha_{s}({\rm 1GeV})}{\alpha_{s}(\mu)}\right]^{\frac{12}{33-2n_f}}\, , \nonumber\\
 \langle\bar{q}g_s \sigma Gq \rangle(\mu)&=&\langle\bar{q}g_s \sigma Gq \rangle({\rm 1GeV})\left[\frac{\alpha_{s}({\rm 1GeV})}{\alpha_{s}(\mu)}\right]^{\frac{2}{33-2n_f}}\, , \nonumber\\
 m_c(\mu)&=&m_c(m_c)\left[\frac{\alpha_{s}(\mu)}{\alpha_{s}(m_c)}\right]^{\frac{12}{33-2n_f}} \, ,\nonumber\\
 m_q(\mu)&=&m_q({\rm 2GeV})\left[\frac{\alpha_{s}(\mu)}{\alpha_{s}({\rm 2GeV})}\right]^{\frac{12}{33-2n_f}} \, ,\nonumber\\
\alpha_s(\mu)&=&\frac{1}{b_0t}\left[1-\frac{b_1}{b_0^2}\frac{\log t}{t} +\frac{b_1^2(\log^2{t}-\log{t}-1)+b_0b_2}{b_0^4t^2}\right]\, ,
\end{eqnarray}
 from the renormalization group equation,  where $q=u$, $d$, $s$,  $t=\log \frac{\mu^2}{\Lambda_{QCD}^2}$, $b_0=\frac{33-2n_f}{12\pi}$, $b_1=\frac{153-19n_f}{24\pi^2}$, $b_2=\frac{2857-\frac{5033}{9}n_f+\frac{325}{27}n_f^2}{128\pi^3}$,  $\Lambda_{QCD}=210\,\rm{MeV}$, $292\,\rm{MeV}$  and  $332\,\rm{MeV}$ for the flavors  $n_f=5$, $4$ and $3$, respectively  \cite{PDG,Narison-mix}, then try to get the ideal energy scales.

 In this work, we explore the properties of  the hidden-charm tetraquark states without strange, with strange and with hidden-strange,  it is better to adopt the flavor number $n_f=4$.

 At the initial  points, we take  the standard values of the vacuum condensates $\langle
\bar{q}q \rangle=-(0.24\pm 0.01\, \rm{GeV})^3$, $\langle\bar{s}s\rangle=(0.8\pm0.1)\langle\bar{q}q\rangle$,
$\langle\bar{q}g_s\sigma G q \rangle=m_0^2\langle \bar{q}q \rangle$,
$\langle\bar{s}g_s\sigma G s \rangle=m_0^2\langle \bar{s}s \rangle$,
$m_0^2=(0.8 \pm 0.1)\,\rm{GeV}^2$,  $\langle \frac{\alpha_s
GG}{\pi}\rangle=(0.012\pm0.004)\,\rm{GeV}^4 $    at the   energy scale  $\mu=1\, \rm{GeV}$
\cite{SVZ79-1,SVZ79-2,Reinders85,Colangelo-Review}, where $q=u$, $d$,  and take the $\overline{MS}$ quark masses $m_{c}(m_c)=(1.275\pm0.025)\,\rm{GeV}$ and $m_s(\mu=2\,\rm{GeV})=(0.095\pm0.005)\,\rm{GeV}$ from the Particle Data Group \cite{PDG}. In numerical computations, we neglect the small masses of the $u$ and $d$ quarks.

In this work, we resort to our unique benchmark, the energy scale formula $\mu=\sqrt{M^2_{X/Y/Z}-(2{\mathbb{M}}_c)^2}$ with the effective $c$-quark mass ${\mathbb{M}}_c=1.82\,\rm{GeV}$ to get  the best energy scales characterizing the QCD spectral densities \cite{Wang-tetra-formula,WangEPJC-1601-Mc}. For detailed discussions about the energy scale formula, one  can consult Refs.\cite{WZG-HT-EPJC-2891,WZG-EPJC-2963,WZG-EPJC-4274,WZG-X4685}.
We can rewrite the energy scale formula in the following  form,
\begin{eqnarray}\label{formula-Regge}
M^2_{X/Y/Z}&=&\mu^2+C\, ,
\end{eqnarray}
where the constants $C$ have the universal value $4{\mathbb{M}}_c^2$ and are fitted numerically via the QCD sum rules, the tetraquark  masses and the best energy scales characterizing the QCD spectral densities have a  Regge-trajectory-like relation.

We always  refer to the experimental data on  the energy  gaps between the ground states (1S) and first radial excited states (2S) as references  to get  the continuum threshold parameters. Considering for  the possible quantum numbers, decay modes and energy  gaps, if we prefer the assignments in terms of compact  tetraquark states in stead of other assignments, we can tentatively assign  the $Z_c(3900)$, $X(3915)$,  $Z_c(4020)$, $X(4140)$, $X(4500)$,  $Z_c(4430)$, $Z_c(4600)$ and $X(4685)$ as  the  hidden-charm tetraquark states  in perfect union, see Table \ref{1S2S}. From the Table, we obtain the energy  gaps $0.57\sim 0.59 \,\rm{GeV}$ between the 1S and 2S hidden-charm tetraquark states, therefore  we  set the continuum threshold parameters to be  $\sqrt{s_0}=M_Z+0.4\sim0.6\,\rm{GeV}$.

\begin{table}
\begin{center}
\begin{tabular}{|c|c|c|c|c|c|c|c|}\hline\hline
   $J^{PC}$   & 1S              & 2S            & Mass Gaps     & References \\ \hline
   $1^{++}$   & $X(4140)$       & $X(4685)$     & 566\,\rm{MeV} & \cite{WZG-X4685,WZG-Di-X4140-EPJC}    \\ \hline
   $1^{+-}$   & $Z_c(3900)$     & $Z_c(4430)$   & 591\,\rm{MeV} & \cite{Maiani-II-type,Nielsen-1401,WangZG-Z4430-CTP}      \\ \hline
    $0^{++}$  & $X(3915)$       & $X(4500)$     & 588\,\rm{MeV} & \cite{X4140-tetraquark-Lebed,X3915-X4500-EPJC-WZG}    \\ \hline
    $1^{+-}$  & $Z_c(4020)$     & $Z_c(4600)$   & 576\,\rm{MeV} & \cite{ChenHX-Z4600-A,WangZG-axial-Z4600} \\ \hline   \hline
\end{tabular}
\end{center}
\caption{ The mass gaps between the 1S and 2S hidden-charm tetraquark states with the possible assignments. }\label{1S2S}
\end{table}

The pole dominance  and convergence of the operator product expansion  are two elementary   criteria, we should satisfy them to reach reliable QCD sum rules.
Now we define the  pole contributions (PC),
\begin{eqnarray}
{\rm{PC}}&=&\frac{\int_{4m_{c}^{2}}^{s_{0}}ds\rho_{QCD}\left(s\right)\exp\left(-\frac{s}{T^{2}}\right)} {\int_{4m_{c}^{2}}^{\infty}ds\rho_{QCD}\left(s\right)\exp\left(-\frac{s}{T^{2}}\right)}\, ,
\end{eqnarray}
 and the contributions of the terms involving the vacuum condensates  of dimension $n$,
\begin{eqnarray}
D(n)&=&\frac{\int_{4m_{c}^{2}}^{s_{0}}ds\rho_{QCD,n}(s)\exp\left(-\frac{s}{T^{2}}\right)}
{\int_{4m_{c}^{2}}^{s_{0}}ds\rho_{QCD}\left(s\right)\exp\left(-\frac{s}{T^{2}}\right)}\, ,
\end{eqnarray}
where the total contributions are normalized  to be 1.

\begin{figure}
 \centering
 \includegraphics[totalheight=8cm,width=10cm]{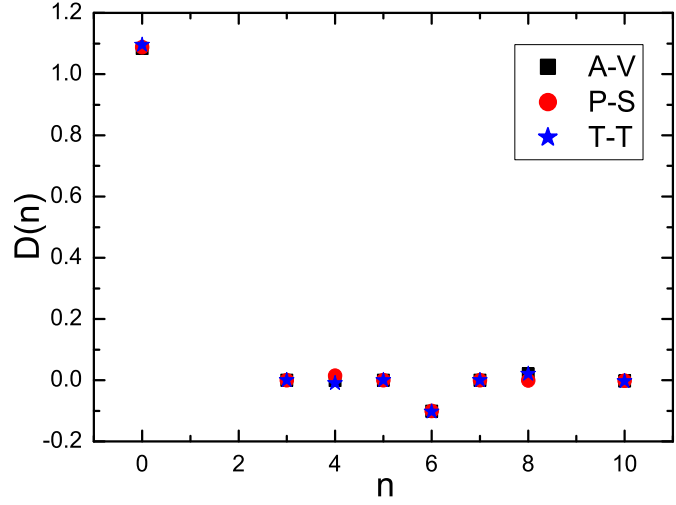}
  \caption{ The contributions of the vacuum condensates with the central values of the input parameters for the tetraquark states with the $J^{PC}=0^{-+}$, where
  the $A-V$, $P-S$ and $T-T$ denote the   $[uc]_{A}[\bar{d}\bar{c}]_{V}-[uc]_{V}[\bar{d}\bar{c}]_{A}$,
$[uc]_{P}[\bar{d}\bar{c}]_{S}+[uc]_{S}[\bar{d}\bar{c}]_{P}$ and  $[uc]_{T}[\bar{d}\bar{c}]_{T}+[uc]_{T}[\bar{d}\bar{c}]_{T}$ tetraquark states,  respectively.   }\label{OPE-fig}
\end{figure}

 We search for the best Borel parameters and continuum threshold parameters in the framework of  trial and error following the routine in our previous works \cite{WZG-HC-spectrum-PRD,WZG-Zcs-spectrum-CPC,WZG-Vector-spectrum-NPB,WZG-Axial-cc-APPB,WZG-cc-spectrum-EPJC,Vector-Tetra-WZG-P-wave-1,Vector-Tetra-WZG-P-wave,WZG-mole-spectrum-IJMPA}. At last,  we reach the satisfactory destinations, such as  the Borel windows, continuum threshold parameters, energy scales of the QCD spectral densities and  pole contributions, which are shown distinctly   in Table \ref{BorelP}.
From the Table,  we can see distinctly that the pole contributions are about $(40-60)\%$ at the hadron side, just like in our previous works  investigating  the hidden-charm (doubly-charmed) tetraquark states with the $J^{PC}=0^{++}$, $1^{++}$, $1^{+-}$, $1^{--}$, $1^{-+}$ and $2^{++}$ ($0^{++}$, $1^{++}$, $1^{+-}$ and $2^{++}$) \cite{WZG-HC-spectrum-PRD,WZG-Zcs-spectrum-CPC,WZG-Vector-spectrum-NPB,WZG-Axial-cc-APPB,WZG-cc-spectrum-EPJC,WZG-mole-spectrum-IJMPA},
 while the central values are larger than $50\%$, the pole dominance criterion  is  matched  with very good.
As an example, in Fig.\ref{OPE-fig}, we plot the contributions of the vacuum condensates $D(n)$ under the condition of the central values of all the input parameters
for the $[uc]_{A}[\bar{d}\bar{c}]_{V}-[uc]_{V}[\bar{d}\bar{c}]_{A}$,
$[uc]_{P}[\bar{d}\bar{c}]_{S}+[uc]_{S}[\bar{d}\bar{c}]_{P}$ and  $[uc]_{T}[\bar{d}\bar{c}]_{T}+[uc]_{T}[\bar{d}\bar{c}]_{T}$ tetraquark states with the $J^{PC}=0^{-+}$. The figure displays that the dominant contributions come  from  the perturbative terms, compared with the lower vacuum condensates, the higher  vacuum condensates play a minor (or tiny) important role (or have very little effects), especially  $|D(10)|\ll 1\%$.

\begin{figure}
 \centering
 \includegraphics[totalheight=6cm,width=7cm]{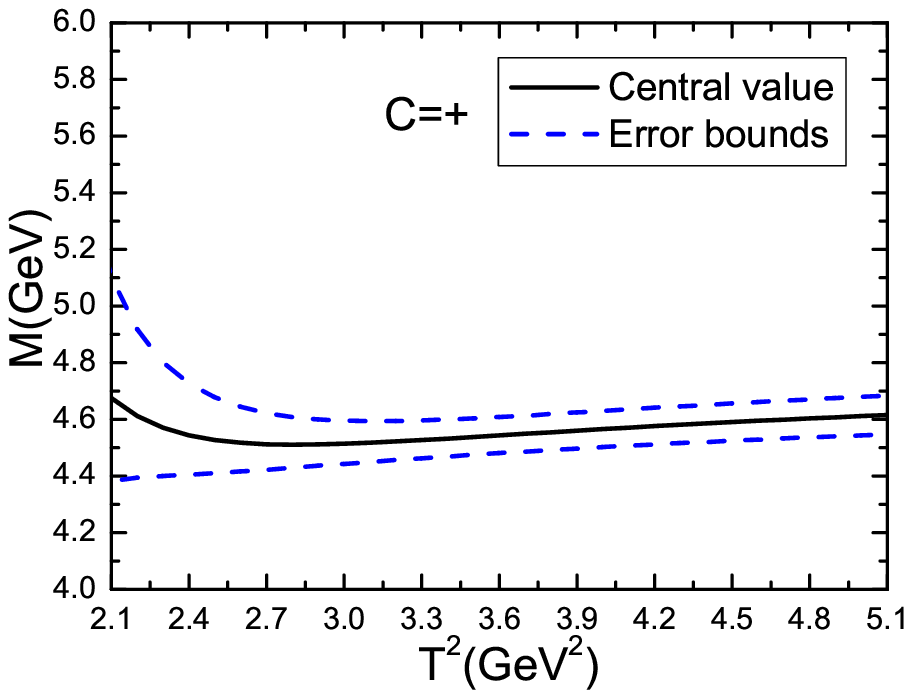}
 \includegraphics[totalheight=6cm,width=7cm]{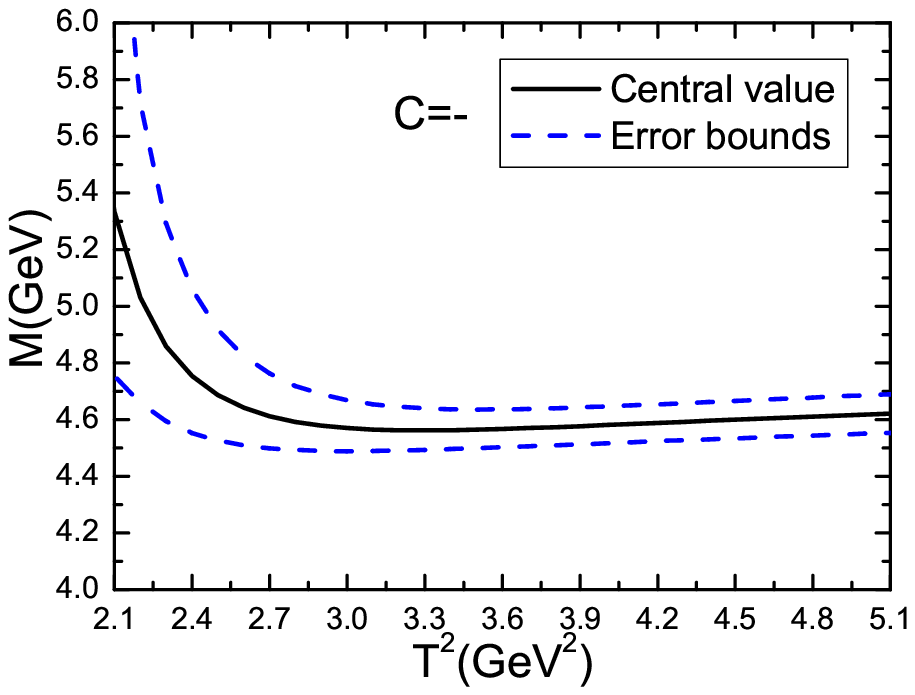}
  \caption{ The masses of  the pseudoscalar  tetraquark states $[uc]_{A}[\bar{d}\bar{c}]_{V}-[uc]_{V}[\bar{d}\bar{c}]_{A}$ ($C=+$) and $[uc]_{A}[\bar{d}\bar{c}]_{V}+[uc]_{V}[\bar{d}\bar{c}]_{A}$ ($C=-$) with variations of the Borel parameters.   }\label{mass-AV}
\end{figure}

We take  all the uncertainties of the input parameters into consideration and acquire  the masses and pole residues of the pseudoscalar hidden-charm  tetraquark states without strange, with strange and with hidden-strange having  the quantum numbers $J^{PC}=0^{-+}$ and $0^{--}$, and we also present them distinctly in Table \ref{BorelP}. From  Table \ref{BorelP}, we can see distinctly that the modified energy scale formula $\mu=\sqrt{M^2_{X/Y/Z}-(2{\mathbb{M}}_c)^2}-k\, m_s(\mu)$ with $k=0$, $1$ or $2$ is  strikingly satisfied, where we subtract the small $s$-quark mass approximately to account for the small light flavor $SU(3)$ mass-breaking effects. In calculations, we have added an uncertainty $\delta\mu=\pm 0.1\,\rm{GeV}$ to the energy scales $\mu$ to
account for the possible uncertainty in determining the effective $c$-quark mass  ${\mathbb{M}}_c$.
 In  Fig.\ref{mass-AV}, we plot the predicted masses of the  $[uc]_{A}[\bar{d}\bar{c}]_{V}-[uc]_{V}[\bar{d}\bar{c}]_{A}$ and $[uc]_{A}[\bar{d}\bar{c}]_{V}+[uc]_{V}[\bar{d}\bar{c}]_{A}$  tetraquark states with the  $J^{PC}=0^{-+}$ and $0^{--}$ respectively via  the variations of the Borel parameters at much larger ranges than the Borel widows as a typical example. From Fig.\ref{mass-AV} and Table \ref{BorelP}, we can see distinctly  that the values of the tetraquark  masses  emerge  smoothly flat platforms in the Borel windows, which can lead to  reliable  predictions.

As can be seen distinctly from Table \ref{mass-residue} that the lowest mass of the pseudoscalar hidden-charm tetraquark state with the symbolic quark constituents  $c\bar{c}u\bar{d}$ is about $4.56\pm0.08\,\rm{GeV}$, which is much larger than the value   $4239\pm18{}^{+45}_{-10}\,\rm{MeV}$ from the LHCb collaboration \cite{LHCb-Zc4430-Rc4240}.

In the dynamical quark model, the lowest masses of the pseudoscalar hidden-charm tetraquark states with the symbolic quark constituents  $c\bar{c}u\bar{d}$ are about $4.2\sim 4.3\,\rm{GeV}$ \cite{Lebed-PRD-2020}, while in the constituent quark models, the lowest masses of the pseudoscalar hidden-charm tetraquark states are about $4.25\,\rm{GeV}$ \cite{Ferretti-PRD-2018}, $4.55\sim 4.60\,\rm{GeV}$ \cite{QFLu-PRD-2016}. In Refs.\cite{Lebed-PRD-2020,Ferretti-PRD-2018,QFLu-PRD-2016}, the authors prefer the explicit P-waves, which  lie between the diquark and antidiquark constituents. The present predictions are compatible with the calculations in Ref.\cite{QFLu-PRD-2016}, but we should bear in mind that the P-waves are implicitly embodied in the negative parity of the diquarks (or antidiquarks) themselves  in the present work, which differ from the quark structures in Refs.\cite{Lebed-PRD-2020,Ferretti-PRD-2018,QFLu-PRD-2016} remarkably.

In Ref.\cite{Chen-Zhu-PRD-2010},  Chen and Zhu study the hidden-charm tetraquark states with the symbolic quark constituents  $c\bar{c}u\bar{d}$ in the framework of  the QCD sum rules, and  obtain the ground state  masses $4.55\pm0.11\,\rm{GeV}$ for the tetraquark states with the $J^{PC}=0^{--}$, the masses $4.55\pm0.11\,\rm{GeV}$, $4.67\pm0.10\,\rm{GeV}$, $4.72\pm0.10\,\rm{GeV}$
 for the tetraquark states with the $J^{PC}=0^{-+}$. The present predictions are consistent with their calculations, again, we should bear in mind that their interpolating currents and  schemes  in treating the operator product expansion and input parameters at the QCD side   differ from the present work remarkably.
 Any current operator  with the same quantum numbers and same quark structure as a Fock state in a hadron couples potentially to this  hadron, so we can construct several current operators to interpolate a hadron, or construct a current operator to interpolate several hadrons. The compare  between the present work and Ref.\cite{Chen-Zhu-PRD-2010} is not entirely vague at all.

From Table \ref{mass-residue}, we can see distinctly that the central values of the masses of the $J^{PC}=0^{-+}$ tetraquark states   with the symbolic quark constituents $uc\bar{d}\bar{c}$, $uc\bar{s}\bar{c}$, $sc\bar{s}\bar{c}$  are about $4.56\sim4.58\,\rm{GeV}$,  $4.61\sim4.62\,\rm{GeV}$ and $4.66\sim4.67\,\rm{GeV}$, respectively,
the central values of the masses of the $J^{PC}=0^{--}$ tetraquark states with the symbolic quark constituents $uc\bar{d}\bar{c}$, $uc\bar{s}\bar{c}$ and $sc\bar{s}\bar{c}$  are about $4.58\,\rm{GeV}$,  $4.63\,\rm{GeV}$  and $4.67\,\rm{GeV}$, respectively.
We can obtain the conclusion tentatively that the currents $J^{+}_{AV}(x)$, $J_{PS}^{+}(x)$ and $J_{TT}^{+}(x)$ couple potentially to three different pseudoscalar tetraquark states with almost degenerated masses,  or to one pseudoscalar tetraquark state with three different Fock components;  the currents $J^{-}_{AV}(x)$ and $J_{PS}^{-}(x)$
couple potentially to two different  pseudoscalar tetraquark states with almost degenerated masses,  or to one pseudoscalar tetraquark state with two different Fock components.
As the currents with the same quantum numbers couple potentially to the pseudoscalar tetraquark  states with almost degenerated masses, the mixing effects cannot improve the predictions remarkably if only the tetraquark masses are concerned. All in all, we obtain reasonable predictions for the masses of the pseudoscalar tetraquark states without strange, with strange and with hidden-strange, the central values  are about $4.56\sim4.58\,\rm{GeV}$,  $4.61\sim4.63\,\rm{GeV}$ and $4.66\sim4.67\,\rm{GeV}$, respectively.

 The following two-body strong decays of the pseudoscalar hidden-charm tetraquark states,
 \begin{eqnarray}\label{Two-Body-Decay}
  Z_c(0^{--}) &\to&  \chi_{c1}\rho \, ,\, \eta_c\rho \, ,\,J/\psi a_1(1260)\, ,\, J/\psi \pi \, ,\,   D\bar{D}_0+h.c.\, ,\,  D^*\bar{D}_1+h.c.\, ,\,  D^*\bar{D}+h.c.\, , \nonumber\\
  Z_c(0^{-+}) &\to&  \chi_{c0}\pi \, ,\, \eta_c f_0(500)\, ,\,J/\psi \rho \, ,\,   D\bar{D}_0+h.c.\, ,\,  D^*\bar{D}_1+h.c.\, ,\,  D^*\bar{D}+h.c.\, , \nonumber\\
  Z_{cs}(0^{--}) &\to&  \chi_{c1}K^* \, ,\, \eta_cK^* \, ,\,J/\psi K_1\, ,\, J/\psi K \, ,\,   D_s\bar{D}_0+h.c.\, ,\, D\bar{D}_{s0}+h.c.\, ,\, D_s^*\bar{D}_1+h.c.\, ,\,\nonumber\\
  && D^*\bar{D}_{s1}+h.c.\, ,\, D_s^*\bar{D}+h.c.\, ,\, D^*\bar{D}_s+h.c.\, , \nonumber\\
  Z_{cs}(0^{-+}) &\to&  \chi_{c0}K \, ,\, \eta_c K^*_0(700) \, ,\,J/\psi K^*\, ,\,   D_s\bar{D}_0+h.c.\, ,\, D\bar{D}_{s0}+h.c.\, ,\, D_s^*\bar{D}_1+h.c.\, ,\,\nonumber\\
  && D^*\bar{D}_{s1}+h.c.\, ,\, D_s^*\bar{D}+h.c.\, ,\, D^*\bar{D}_s+h.c.\, ,\nonumber\\
  Z_{css}(0^{--}) &\to&  \chi_{c1}\phi \, ,\, \eta_c\phi \, ,\,J/\psi f_1\, ,\, J/\psi \eta \, ,\,   D_s\bar{D}_{s0}+h.c.\, ,\,  D_s^*\bar{D}_{s1}+h.c.\, ,\,  D_s^*\bar{D}_s+h.c.\, , \nonumber\\
  Z_{css}(0^{-+}) &\to&  \chi_{c0}\eta \, ,\, \eta_c f_0(980)\, ,\,J/\psi \phi \, ,\,    D_s\bar{D}_{s0}+h.c.\, ,\,  D_s^*\bar{D}_{s1}+h.c.\, ,\,  D_s^*\bar{D}_s+h.c.\, ,
\end{eqnarray}
can take place through the Okubo-Zweig-Iizuka super-allowed fall-apart mechanism.   We can  probe those pseudoscalar  hidden-charm  tetraquark  states
 at the BESIII, LHCb, Belle II,  CEPC, FCC, ILC in the future, and  confront the present  predictions   with  the experimental data to illustrate the nature of the exotic $X$, $Y$ and $Z$ states.

 In Ref.\cite{WZG-ZHJX-Zc3900}, we  assign the $Z_c^\pm(3900)$ as  the diquark-antidiquark  type   tetraquark  state with the quantum numbers $J^{PC}=1^{+-}$, and investigate the hadronic coupling  constants in its two-body strong decays with the QCD sum rules based on the novel analysis, i.e. rigorous current-hadron duality,  and obtain  satisfactory total width to match to the experimental data. The novel analysis has been successfully applied to study the decay widths of the $X(4140)$, $X(4274)$, $Y(4660)$, $P_c(4312)$, etc \cite{WZG-Di-X4140-EPJC,WZG-Y4274,WZG-Y4660-decay,WZG-WHJ-Pc4212}. We can extend our previous works to explore the two-body strong decays shown in Eq.\eqref{Two-Body-Decay} with the three-point QCD sum rules based on the rigorous current-hadron duality, and get  the branching fractions, which can be confronted with  the experimental data in the future to identify the pseudoscalar hidden-charm tetraquark states in more reasonable foundations. We prefer to accomplish the complex and arduous calculations in an independent work.

\begin{table}
\begin{center}
\begin{tabular}{|c|c|c|c|c|c|c|c|c|}\hline\hline
 $Z_c$                                                        &$J^{PC}$  &$T^2(\rm{GeV}^2)$ &$\sqrt{s_0}(\rm GeV) $ &$\mu(\rm{GeV})$  &pole     \\ \hline

$[uc]_{A}[\bar{d}\bar{c}]_{V}-[uc]_{V}[\bar{d}\bar{c}]_{A}$   &$0^{-+}$  &$3.7-4.1$          &$5.10\pm0.10$          &$2.7$           &$(42-60)\%$ \\ \hline

$[uc]_{A}[\bar{d}\bar{c}]_{V}+[uc]_{V}[\bar{d}\bar{c}]_{A}$   &$0^{--}$  &$3.7-4.1$          &$5.10\pm0.10$          &$2.8$           &$(42-60)\%$ \\ \hline

$[uc]_{A}[\bar{s}\bar{c}]_{V}-[uc]_{V}[\bar{s}\bar{c}]_{A}$   &$0^{-+}$  &$3.7-4.1$          &$5.15\pm0.10$          &$2.7$           &$(43-61)\%$ \\ \hline

$[uc]_{A}[\bar{s}\bar{c}]_{V}+[uc]_{V}[\bar{s}\bar{c}]_{A}$   &$0^{--}$  &$3.7-4.1$          &$5.15\pm0.10$          &$2.8$           &$(43-61)\%$ \\ \hline

$[sc]_{A}[\bar{s}\bar{c}]_{V}-[sc]_{V}[\bar{s}\bar{c}]_{A}$   &$0^{-+}$  &$3.8-4.2$          &$5.20\pm0.10$          &$2.7$           &$(42-60)\%$ \\ \hline

$[sc]_{A}[\bar{s}\bar{c}]_{V}+[sc]_{V}[\bar{s}\bar{c}]_{A}$   &$0^{--}$  &$3.8-4.2$          &$5.20\pm0.10$          &$2.8$           &$(43-60)\%$ \\ \hline

$[uc]_{P}[\bar{d}\bar{c}]_{S}+[uc]_{S}[\bar{d}\bar{c}]_{P}$   &$0^{-+}$  &$3.7-4.1$          &$5.10\pm0.10$          &$2.8$           &$(42-60)\%$ \\ \hline

$[uc]_{P}[\bar{d}\bar{c}]_{S}-[uc]_{S}[\bar{d}\bar{c}]_{P}$   &$0^{--}$  &$3.7-4.1$          &$5.10\pm0.10$          &$2.8$           &$(42-60)\%$ \\ \hline

$[uc]_{P}[\bar{s}\bar{c}]_{S}+[uc]_{S}[\bar{s}\bar{c}]_{P}$   &$0^{-+}$  &$3.7-4.1$          &$5.15\pm0.10$          &$2.8$           &$(43-61)\%$ \\ \hline

$[uc]_{P}[\bar{s}\bar{c}]_{S}-[uc]_{S}[\bar{s}\bar{c}]_{P}$   &$0^{--}$  &$3.7-4.1$          &$5.15\pm0.10$          &$2.8$           &$(43-61)\%$ \\ \hline

$[sc]_{P}[\bar{s}\bar{c}]_{S}+[sc]_{S}[\bar{s}\bar{c}]_{P}$   &$0^{-+}$  &$3.8-4.2$          &$5.20\pm0.10$          &$2.8$           &$(43-61)\%$ \\ \hline

$[sc]_{P}[\bar{s}\bar{c}]_{S}-[sc]_{S}[\bar{s}\bar{c}]_{P}$   &$0^{--}$  &$3.8-4.2$          &$5.20\pm0.10$          &$2.8$           &$(43-61)\%$ \\ \hline

$[uc]_{T}[\bar{d}\bar{c}]_{T}+[uc]_{T}[\bar{d}\bar{c}]_{T}$   &$0^{-+}$  &$3.7-4.1$          &$5.10\pm0.10$          &$2.7$           &$(41-60)\%$ \\ \hline

$[uc]_{T}[\bar{s}\bar{c}]_{T}+[uc]_{T}[\bar{s}\bar{c}]_{T}$   &$0^{-+}$  &$3.7-4.1$          &$5.15\pm0.10$          &$2.7$           &$(43-61)\%$ \\ \hline

$[sc]_{T}[\bar{s}\bar{c}]_{T}+[sc]_{T}[\bar{s}\bar{c}]_{T}$   &$0^{-+}$  &$3.8-4.2$          &$5.20\pm0.10$          &$2.7$           &$(42-60)\%$ \\ \hline

\hline\hline
\end{tabular}
\end{center}
\caption{ The Borel parameters, continuum threshold parameters, energy scales of the QCD spectral densities and  pole contributions for the pseudoscalar hidden-charm tetraquark states. }\label{BorelP}
\end{table}

\begin{table}
\begin{center}
\begin{tabular}{|c|c|c|c|c|c|c|c|c|}\hline\hline
 $Z_c$                                                        &$J^{PC}$  &$M_Z(\rm{GeV})$   &$\lambda_Z(\rm GeV^5) $     \\ \hline

$[uc]_{A}[\bar{d}\bar{c}]_{V}-[uc]_{V}[\bar{d}\bar{c}]_{A}$   &$0^{-+}$  &$4.56\pm0.08$     &$(1.33\pm0.18)\times 10^{-1}$      \\   \hline

$[uc]_{A}[\bar{d}\bar{c}]_{V}+[uc]_{V}[\bar{d}\bar{c}]_{A}$   &$0^{--}$  &$4.58\pm0.07$     &$(1.37\pm0.17)\times 10^{-1}$      \\   \hline

$[uc]_{A}[\bar{s}\bar{c}]_{V}-[uc]_{V}[\bar{s}\bar{c}]_{A}$   &$0^{-+}$  &$4.61\pm0.08$     &$(1.41\pm0.19)\times 10^{-1}$      \\   \hline

$[uc]_{A}[\bar{s}\bar{c}]_{V}+[uc]_{V}[\bar{s}\bar{c}]_{A}$   &$0^{--}$  &$4.63\pm0.08$     &$(1.45\pm0.19)\times 10^{-1}$      \\   \hline

$[sc]_{A}[\bar{s}\bar{c}]_{V}-[sc]_{V}[\bar{s}\bar{c}]_{A}$   &$0^{-+}$  &$4.66\pm0.08$     &$(1.50\pm0.20)\times 10^{-1}$      \\   \hline

$[sc]_{A}[\bar{s}\bar{c}]_{V}+[sc]_{V}[\bar{s}\bar{c}]_{A}$   &$0^{--}$  &$4.67\pm0.08$     &$(1.53\pm0.20)\times 10^{-1}$      \\   \hline

$[uc]_{P}[\bar{d}\bar{c}]_{S}+[uc]_{S}[\bar{d}\bar{c}]_{P}$   &$0^{-+}$  &$4.58\pm0.07$     &$(6.92\pm0.86)\times 10^{-2}$      \\   \hline

$[uc]_{P}[\bar{d}\bar{c}]_{S}-[uc]_{S}[\bar{d}\bar{c}]_{P}$   &$0^{--}$  &$4.58\pm0.07$     &$(6.91\pm0.86)\times 10^{-2}$      \\   \hline

$[uc]_{P}[\bar{s}\bar{c}]_{S}+[uc]_{S}[\bar{s}\bar{c}]_{P}$   &$0^{-+}$  &$4.63\pm0.07$     &$(7.30\pm0.90)\times 10^{-2}$      \\   \hline

$[uc]_{P}[\bar{s}\bar{c}]_{S}-[uc]_{S}[\bar{s}\bar{c}]_{P}$   &$0^{--}$  &$4.63\pm0.07$     &$(7.30\pm0.90)\times 10^{-2}$      \\   \hline

$[sc]_{P}[\bar{s}\bar{c}]_{S}+[sc]_{S}[\bar{s}\bar{c}]_{P}$   &$0^{-+}$  &$4.67\pm0.08$     &$(7.73\pm0.97)\times 10^{-2}$      \\   \hline

$[sc]_{P}[\bar{s}\bar{c}]_{S}-[sc]_{S}[\bar{s}\bar{c}]_{P}$   &$0^{--}$  &$4.67\pm0.08$     &$(7.73\pm0.96)\times 10^{-2}$      \\   \hline

$[uc]_{T}[\bar{d}\bar{c}]_{T}+[uc]_{T}[\bar{d}\bar{c}]_{T}$   &$0^{-+}$  &$4.57\pm0.08$     &$(4.62\pm0.61)\times 10^{-1}$      \\   \hline

$[uc]_{T}[\bar{s}\bar{c}]_{T}+[uc]_{T}[\bar{s}\bar{c}]_{T}$   &$0^{-+}$  &$4.62\pm0.08$     &$(4.89\pm0.63)\times 10^{-1}$      \\   \hline

$[sc]_{T}[\bar{s}\bar{c}]_{T}+[sc]_{T}[\bar{s}\bar{c}]_{T}$   &$0^{-+}$  &$4.67\pm0.08$     &$(5.19\pm0.67)\times 10^{-1}$      \\   \hline

\hline\hline
\end{tabular}
\end{center}
\caption{ The masses and pole residues  for the ground state pseudoscalar  hidden-charm tetraquark states. }\label{mass-residue}
\end{table}

\section{Conclusion}
In the present work,  we take all the color-antitriplet diquark operators, such as the scalar, pseudoscalar,  vector,  axialvector and tensor  diquark operators,  as the elementary  building blocks    to construct the  four-quark currents  without importing the explicit P-waves to implement the negative-parity,   and take account of all the light flavor $SU(3)$ breaking effects, such as the vacuum condensates and quark masses, to investigate the mass spectroscopy of the pseudoscalar  hidden-charm tetraquark states without strange, with strange and with hidden-strange in the framework of  the QCD sum rules  comprehensively as a further  extension of our previous works. We obtain the lowest mass $4.56\pm0.08\,\rm{GeV}$ for the tetraquark state with the symbolic quark constituents  $c\bar{c}u\bar{d}$, which is much larger than the experimental value   $4239\pm18{}^{+45}_{-10}\,\rm{MeV}$ from the LHCb collaboration, and the discrepancy does not support assigning the $Z_c(4240)$, which still needs confirmation, to be the pseudoscalar hidden-charm tetraquark state with the  symbolic quark constituents  $c\bar{c}u\bar{d}$.
We can  search for those pseudoscalar  hidden-charm  tetraquark  states at the Okubo-Zweig-Iizuka super-allowed two-body strong decays
 at the BESIII, LHCb, Belle II,  CEPC, FCC, ILC in the future, and  confront the present  predictions   with the experimental data to examine reliability of the calculations.

\section*{Acknowledgements}
This  work is supported by National Natural Science Foundation, Grant Number  12175068.

\end{document}